\documentclass[12pt]{article}
\usepackage{a4}
\usepackage{amsmath}
\usepackage{amsfonts}
\usepackage{amsmath}
\usepackage{latexsym}
\usepackage{amsfonts}
\usepackage{graphics}
\usepackage{graphicx}
\usepackage{epic}
\usepackage{eepic}
\begin{document}
\oddsidemargin 6pt\evensidemargin 6pt\marginparwidth
48pt\marginparsep 10pt

\renewcommand{\thefootnote}{\fnsymbol{footnote}}
\thispagestyle{empty}

\noindent    \hfill  February    2006 \\

\noindent \vskip3.3cm
\begin{center}

{\Large\bf The structure of the trace anomaly  of higher spin
conformal currents in  the bulk of $AdS_{4}$}
\bigskip\bigskip\bigskip

{\large Ruben Manvelyan ${}^{\dag \ddag}$ and Werner R\"uhl
${}^{\dag}$}
\medskip

${}^{\dag}${\small\it Department of Physics\\ Erwin Schr\"odinger Stra\ss e \\
Technical University of Kaiserslautern, Postfach 3049}\\
{\small\it 67653
Kaiserslautern, Germany}\\
\medskip
${}^{\ddag}${\small\it Yerevan Physics Institute\\ Alikhanian Br.
Str.
2, 375036 Yerevan, Armenia}\\
\medskip
{\small\tt manvel,ruehl@physik.uni-kl.de}
\end{center}

\bigskip 
\begin{center}
{\sc Abstract}
\end{center}

The two-point function of the conserved traceless spin-$\ell$
currents which are constructed from the scalar field $\sigma(z)$ is
evaluated and renormalized by a dimensional regularization
procedure. The anomaly is managed to arise only in the trace part.
To isolate this trace anomaly it is sufficient to analyze only the
maximum singular part of the two-point function and its trace terms
to leading order. The corresponding part of the effective action
which is quadratic in the trace of the higher spin field is
explicitly given. For the spin-2 field which is identical with the
gravitational field the results known from the literature are
reproduced.

\newpage

\section{Introduction}
The $AdS_{4}/CFT_{3}$ correspondence of the critical $O(N)$ sigma
model and four dimensional higher spin gauge theory in anti de
Sitter space  \cite{Klebanov} increased interest in the complicated
problem of quantization and interaction of the higher spin gauge
theories in $AdS$ space \cite{Frons, Vasiliev}. In this case the
well known boundary theory has been used for the reconstruction of
the unknown interacting field theory in the bulk of $AdS_{4}$ \cite{
R,Ruehl,T,RM3,RM2}. This $AdS_{4}$ case is interesting also in view
of the presence of the conformal anomaly in even dimensional
space-time \cite{anom}. The anomaly should arise during the one-loop
calculation and can be used for checking of the correctness of the
linearized interaction between higher spin gauge and scalar fields
\cite{RM} and always should reproduce for $\ell=2$ the well known
results of the trace anomaly of the scalar mode in curved background
\cite{BD}. Investigation of this problem could also be important for
a deeper understanding of the geometrical and topological structure
of the linearized interaction of the higher spin gauge fields.

In the previous article \cite{RM4} we investigated the anomalous
behaviour of the renormalized one-loop effective action for the
conformal scalar mode in the external higher spin gauge field
\begin{eqnarray}
&&W(h^{(\ell)})=\frac{1}{2\ell^{2}}\int_{z_{1}}\int_{z_{2}}h^{(\ell)\mu_{1}\dots\mu_{\ell}}
   (z_{1})< J^{(\ell)}_{\mu_{1}\dots\mu_{\ell}}(z_{1}) J^{(\ell)}_{\nu_{1}\dots\nu_{\ell}}(z_{2})
    >h^{(\ell)\nu_{1}\dots\nu_{\ell}}(z_{2}) \label{efac}.  \quad\\
    && \int_{z}:=\int d^{4}z\sqrt{g(z)}
\end{eqnarray}
Here $h^{(\ell)\mu_{1}\dots\mu_{\ell}}(z)$ is the spin $\ell$ gauge
field and $J^{(\ell)}_{\mu_{1}\dots\mu_{\ell}}(z)$ the conserved
traceless current constructed from the conformally coupled scalar
$\sigma(z)$ \cite{RM} in $D=d+1$ dimensional $AdS$ space\footnote{We
will use Euclidian $AdS_{d+1}$ with conformal flat metric, curvature
and covariant derivatives satisfying
\begin{eqnarray}
&&ds^{2}=g_{\mu \nu }(z)dz^{\mu }dz^{\nu
}=\frac{L^{2}}{(z^{0})^{2}}\delta _{\mu \nu }dz^{\mu }dz^{\nu
},\quad \sqrt{g}=\frac{L^{d+1}}{(z^{0})^{d+1}}\;,
\notag  \label{ads} \\
&&\left[ \nabla _{\mu },\,\nabla _{\nu }\right] V_{\lambda }^{\rho
}=R_{\mu \nu \lambda }^{\quad \,\,\sigma }V_{\sigma }^{\rho }-R_{\mu
\nu \sigma
}^{\quad \,\,\rho }V_{\lambda }^{\sigma }\;,  \notag \\
&&R_{\mu \nu \lambda }^{\quad \,\,\rho
}=-\frac{1}{(z^{0})^{2}}\left( \delta _{\mu \lambda }\delta _{\nu
}^{\rho }-\delta _{\nu \lambda }\delta _{\mu }^{\rho }\right)
=-\frac{1}{L^{2}}\left( g_{\mu \lambda }(z)\delta _{\nu
}^{\rho }-g_{\nu \lambda }(z)\delta _{\mu }^{\rho }\right) \;,  \notag \\
&&R_{\mu \nu }=-\frac{d}{(z^{0})^{2}}\delta _{\mu \nu }=-\frac{d}{L^{2}}%
g_{\mu \nu }(z)\quad ,\quad R=-\frac{d(d+1)}{L^{2}}\;.  \notag
\end{eqnarray}%
From now on we put $L=1$.}. The additional factor
$\frac{1}{\ell^{2}}$ comes from the following normalization of the
linearized interaction \setlength{\unitlength}{0.252mm}
\begin{equation}\label{clint}
    S^{(\ell)conf}_{int}=\frac{1}{\ell}\int d^{4}z\sqrt{g(z)}
    h^{(\ell)\mu_{1}\dots\mu_{\ell}}(z)J^{(\ell)}_{\mu_{1}\dots\mu_{\ell}}(z)
\end{equation}
leading to the following rule of calculation of the correlation
functions
\begin{equation}\label{cf}
    <J^{\ell}_{\mu_{1}\dots\mu_{\ell}}(z)\dots>:=\frac{\ell}{\sqrt{g(z)}}\frac{\delta}{\delta
    h^{(\ell)\mu_{1}\dots\mu_{\ell}}(z)}\dots  W(h^{(\ell)}) .
\end{equation}
This is in agreement with the usual normalization of the
energy-momentum tensor ($\ell=2$ case) for the conformally coupled
scalar matter field \cite{BD}
\begin{equation}\label{emt}
    <T_{\mu\nu}(z)>:=\frac{2}{\sqrt{g(z)}}\frac{\delta}{\delta
    g^{\mu\nu}(z)}W(g_{\mu\nu}) .
\end{equation}

In the future for shortening the notation and calculation we
contract all rank $\ell$ symmetric tensor fields at $z$  with the $\ell $-fold
tensor product of a vector $a^{\mu }$ from the tangential space at $z$. In this notation the
symmetric tensor, it's trace, symmetrized gradient and divergence
can be written as
\begin{eqnarray}
&&
J^{(\ell)}(a;z)=J^{(\ell)}_{\mu_{1}\dots\mu_{\ell}}a^{\mu_{1}}\dots
a^{\mu_{\ell}} ,\\
  && TrJ^{(\ell)}(a;z)=\frac{1}{\ell(\ell-1)}\Box_{a}J^{(\ell)}(a;z)
  ,\quad
  \Box_{a}=g^{\mu\nu}(z)\frac{\partial^{2}}{\partial a^{\mu}\partial a^{\nu}} ,\\
  && (a^{\mu}\nabla_{\mu})J^{(\ell)}(a;z) ,\quad  \nabla \cdot
  J^{(\ell)}(a;z)=\frac{1}{\ell}\nabla^{\mu}\frac{\partial}{\partial
  a^{\mu}}J^{(\ell)}(a;z) .
\end{eqnarray}
We introduce also the following notations for contractions in $z$
and $a$ spaces
\begin{eqnarray}
  && \int d^{4}z\sqrt{g(z)}J^{(\ell)}_{\mu_{1}\dots\mu_{\ell}}(z)h^{(\ell)\mu_{1}\dots\mu_{\ell}}(z)=
  J^{(\ell)}(a;z)\ast_{a_{z}}h^{(\ell)}(a;z)\quad , \\
  &&
  J^{(\ell)}_{\mu_{1}\dots\mu_{\ell}}h^{(\ell)\mu_{1}\dots\mu_{\ell}}=J^{(\ell)}(a)\hat{\ast}_{a}h^{(\ell)}(a)
\end{eqnarray}

 In \cite{RM4} we considered the extraction of
singularities and anomalous behaviour of the two point function $<
J^{(\ell)}(z_{1};a) J^{(\ell)}(z_{2};c)
>$ due to the "trace terms". It means that we performed all our
calculations with the exception of $O(a^{2})$ and $O(c^{2})$ terms.
This was enough for the investigation of the renormalized Ward
identities and the observation of the violation of the tracelessness
condition under the stipulation that the renormalized two-point
function satisfies the Ward identities following from the
conservation condition (gauge invariance). In this article we
enlarge our consideration and include the first $O(a^{2})$ and
$O(c^{2})$ terms in order to understand the \emph{exact structure
and anomaly coefficients of the conformal scalar mode in the
external higher spin field}. To control our calculation we will
compare the particular case $\ell=2$ with the known results for the
trace anomaly in an external gravitational field \cite{BD} and the
connection of the anomalous coefficients with the correlation
functions in the flat \cite{OP} and the constant curvature
backgrounds \cite{Osborn} .

\section{Loop function and Ward identity with trace terms}
In the previous article we calculated the one-loop two-point
function
\begin{equation}\label{lf}
  \Pi^{(\ell)}(z_{1};a|z_{2};c):=  < J^{(\ell)}(z_{1};a) J^{(\ell)}(z_{2};c) >
\end{equation}
applying just Wick's theorem to the leading part of the conserved
and traceless currents avoiding $O(a^{2})$ and $O(c^{2})$
contributions from so-called "trace terms" \cite{RM4}. Calculations
were done using the propagator of the scalar field in $AdS_{4}$
quantized with a boundary condition corresponding to the free
conformal point of the boundary $O(N)$ model
\begin{equation}\label{prs}
<\sigma(z_{1})
\sigma(z_{2})>=\frac{1}{8\pi^{2}}\left(\frac{1}{u}+\frac{1}{u+2}\right)
,
\end{equation}
where
\begin{equation}\label{zeta}
    u=\zeta-1=\frac{(\vec{z}_{1}-\vec{z}_{2})^{2}}{2z^{0}_{1}z^{0}_{2}}
\end{equation}
is the invariant chordal  distance in $AdS$.

The general rule for working with such objects is analyzed in detail
in the same article. The main point is the following. The tensorial
structure of any two-point function in $AdS$ space can be described
using a general  basis  of the independent bitensors
\cite{AllenJ,AllenT,Turyn,LMR1,LMR2,Freed}
     \begin{eqnarray}
       && I_{1}(a,c):=(a\partial)_{1}(c\partial)_{2}u(z_{1},z_{2}) , \\
       && I_{2}(a,c):=(a\partial)_{1}u(z_{1},z_{2})(c\partial)_{2}u(z_{1},z_{2}),\\
       && I_{3}(a,c):=a^{2}_{1}I^{2}_{2c}+c^{2}_{2}I^{2}_{1a} , \\
       && I_{4}:=a^{2}_{1}c^{2}_{2} ,\\
       && I_{1a}:=(a\partial)_{1}u(z_{1},z_{2})\quad ,
       \quad I_{2c}:=(c\partial)_{2}u(z_{1},z_{2}) , \\
       &&(a\partial)_{1}=a^{\mu}\frac{\partial}{\partial
       z_{1}^{\mu}} ,\quad (c\partial)_{2}=c^{\mu}\frac{\partial}{\partial
       z_{2}^{\mu}} ,\\&& a^{2}_{1}=g_{\mu\nu}(z_{1})a^{\mu}a^{\nu} ,
       \quad c^{2}_{2}=g_{\mu\nu}(z_{2})c^{\mu} c^{\nu} .
     \end{eqnarray}
In this case this basis should appear automatically after
contractions of scalars and action of the vertex  derivatives. In
general we have to get an expansion with all four basis elements
\begin{equation}\label{K}
    \Pi^{\ell}(z_{1};a|z_{2};c)=\Psi^{\ell}[F]+\sum_{n,m;\, 0<2(n+m)<\ell}I_{3}^{n}I_{4}^{m}
    \Psi^{\ell-2(n+m)}[G^{(n,m)}] .
\end{equation}
Here we introduce a special map from the set
$\{F_{k}(u)\}^{\ell}_{k=0}$ of the $\ell+1$ functions of $u$ to the
space of $\ell\times \ell$ bitensors
\begin{equation}\label{Psy}
    \Psi^{\ell}[F]=\sum^{\ell}_{k=0}I^{\ell-k}_{1}(a,c)I^{k}_{2}(a,c)F_{k}(u)
    .\nonumber
\end{equation}

But in \cite{RM4} we restricted our consideration on the first part
of (\ref{K}) connected with the $I_{1}, I_{2}$ bitensors only
calling all monomials corresponding to $I_{3}$ and $I_{4}$ in the
above sum and the corresponding sets of functions
$\{G^{(n,m)}_{k}\}_{k=0}^{\ell-2(n+m)}$ the "trace terms"
\begin{eqnarray}
  \Pi^{\ell}(z_{1};a|z_{2};c)=\Psi^{\ell}[F]+ \textnormal{trace terms} .\label{sets}
\end{eqnarray}
Here for the investigation of the exact structure (not only
existence) of the trace anomaly we take the first   $O(a^{2})$ and
$O(c^{2})$ terms into account. It means that we will use instead of
(\ref{sets}) the following ansatz
\begin{eqnarray}
  && \Pi^{\ell}(z_{1};a|z_{2};c)=n(\ell)K^{\ell}(F,G,H)+\textnormal{next trace terms}
  ,\label{sets1}\\
&& K^{\ell}(F,G,H)=\Psi^{\ell}[F]+ I_{3}\Psi^{\ell-2}[G]+
  I_{4}\Psi^{\ell-2}[H] .\label{aan}
\end{eqnarray}
The normalization factor $n(\ell)$ will be fixed below. This ansatz
must satisfy the following naive Ward identities \footnote{For the
purpose of derivation with respect to $z_{1}$ or $z_{2}$ we
understand the invariant functions $F, G, H$ as analytic functions
on the half plane $\Re u > 0$. Distributions on the positive real
$u$ axis are obtained as boundary values of these later on.}
following from the conservation and tracelessness conditions of our
currents with contribution of the corresponding partner trace terms
described by the expansion in the other two bitensors $I_{3},I_{4}$
\begin{eqnarray}
&&\Box_{a}K^{\ell}(F,G,H)=\frac{\partial^{2}}
    {\partial a_{\mu}\partial a^{\mu}}K^{\ell}(F,G,H)=0 ,\label{Ward1}\\
&&(\nabla\cdot\partial_{a})
K^{\ell}(F,G,H)=\nabla^{\mu}\frac{\partial}{\partial
a^{\mu}}K^{\ell}(F,G,H)=0 .\label{Ward2}
\end{eqnarray} All important
relations for the calculations including $O(a^{2})$ and $O(c^{2})$
terms to be performed below can be found in Appendix A of this
article. In the main text we will only present the Ward identities
(\ref{Ward1}),(\ref{Ward2}) in the language of the sets of functions
$\{F_{k}(u)\}^{\ell}_{k=0}$, $\{G_{k}(u)\}^{\ell-2}_{k=0}$ and
$\{H_{k}(u)\}^{\ell-2}_{k=0}$. For that purpose we apply
(\ref{start})-(\ref{end}) to our ansatz (\ref{aan}) and keep only
the leading trace terms. We get

 1) the Divergence map
\begin{eqnarray}
  \nabla^{\mu}_{1}\frac{\partial}{\partial
  a^{\mu}}K^{\ell}(F,G,H)&=&
  I_{2c}\Psi^{\ell-1}[D^{1}_{\ell}(F,G)]
  +c^{2}_{2}I_{1a}\Psi^{\ell-2}[D^{2}_{\ell}(F,G,H)], \quad\label{div}\\
  D^{1}_{\ell}(F,G)&=&(Div_{\ell}F)_{k}+2(k+2)G_{k}+2G'_{k-1} ,\\
  (Div_{\ell}F)_{k}&=&(\ell-k)(u+1)
  F'_{k}+(k+1)u(u+2)F'_{k+1}\nonumber\\
  &+&(\ell-k)(\ell+d+k)F_{k}+(k+1)(\ell+d+k+1)u F_{k+1}
  ,\label{dv}\quad \quad\\
  D^{2}_{\ell}(F,G,H)&=&(k+1)(\ell-k-1)F_{k+1}+u(u+2)(k+2)G'_{k}\nonumber\\
  &+&(u+1)(\ell-k-1)G'_{k-1}+(u+1)(k+2)(d+\ell+k)G_{k}\nonumber\\
  &+&(\ell-k-1)(\ell+d+k-1)G_{k-1}+2H'_{k}+2(k+1)H_{k+1} ,\quad\quad\label{dv1}
  \end{eqnarray}
  where $ F'_{k}:=\frac{\partial}{\partial
  u}F_{k}(u)$ and so on;

2) the Trace map
\begin{eqnarray}
    \Box_{a}K^{\ell}(F,G,H)&=&I^{2}_{2c}\Psi^{\ell-2}
    [T^{1}_{\ell}(F,G)]+ c^{2}_{2}\Psi^{\ell-2}[T^{2}_{\ell}(F,G,H)] ,\label{tracemap}\\
T^{1}_{\ell}(F,G)&=&(Tr_{\ell}F)_{k}+2(d+2\ell-3)G_{k} ,
\\(Tr_{\ell}F)_{k}&=&(\ell-k)(\ell-k-1)F_{k} +2(k+1)(\ell-k-1)(u+1)
F_{k+1}\nonumber\\&+&(k+2)(k+1)u(u+2)F_{k+2}
,\label{tr}\\
T^{2}_{\ell}(F,G,H)&=&(\ell-k)(\ell-k-1)F_{k}+(k+1)(k+2)u(u+2)G_{k}\nonumber\\
&+& 2(k+1)(\ell-k-1)(u+1)G_{k-1}+(\ell-k)(\ell-k-1)G_{k-2}\nonumber\\
&+& 2(d+2\ell-3)H_{k} .\label{tracemap1}
\end{eqnarray}

Then we obtain two relations from the tracelessness condition
\begin{eqnarray}
  && T^{1}_{\ell}(F,G)=0 ,\label{tw1} \\
  && T^{2}_{\ell}(F,G,H)=0 ,\label{tw2}
\end{eqnarray}
and further two relations from the conservation condition
\begin{eqnarray}
&& D^{1}_{\ell}(F,G)=0 ,\label{d0}\\
  &&D^{2}_{\ell}(F,G,H)=0 .\label{d}
  \end{eqnarray}

In \cite{RM4} we discovered that essential for the renormalization
procedure of (\ref{sets1}) described by the function set
$\{F_{k}(u)\}^{\ell}_{k=0}$ is the \emph{main singular part} of the
form
\begin{eqnarray}\label{dist}
    &&\Pi^{\ell}_{B}(z_{1};a|z_{2};c)=\frac{(-1)^{\ell}(2\ell)!}{2^{7}\pi^{4}
}\sum^{\ell}_{k=0}
 I_{1}^{\ell-k}I_{2}^{k}F_{k}^{B}(u) ,\\&&F_{k}^{B}(u)=(-1)^{k}
 \binom{\ell}{k}\frac{1}{u^{\ell+k+2}} .\label{bare}
\end{eqnarray}
This was proven by direct calculation of the loop diagram after
extraction of the gauge-gradient and regular terms. The next
important point is that this main singular part satisfies the
following simple Ward identities without entanglement of the sets
$\{G_{k}(u)\}^{\ell-2}_{k=0}$ and $\{H_{k}(u)\}^{\ell-2}_{k=0}$
\begin{eqnarray}
  && (Div_{\ell}F^{B})_{k}^{d=3}=0,\quad(Tr_{\ell}F^{B})^{d=3}_{k}=0
  .
\end{eqnarray}
As a result we could, using only the $F$-set, calculate the
corresponding singular part of the effective action (see Eq.(89) of
\cite{RM4}) and prove that after renormalization we  observe a
\emph{trace anomaly}.

For this consideration this result will simplify our tasks in the
following two points:
\begin{itemize}
  \item We can now fix the normalization coefficient $n(\ell)$ in (\ref{sets1})
  \begin{equation}\label{n}
    n(\ell)=\frac{(-1)^{\ell}(2\ell)!}{2^{7}\pi^{4}} ,
  \end{equation}

  \item We can obtain the main singular  $G^{B}$- and $H^{B}$-dependent
  "bare" part of $K^{\ell}(F^{B},G^{B},H^{B})$ without loop
  calculation but just using the Ward identities
  (\ref{tw1})-(\ref{d}) (for d=3)
  and expression (\ref{bare}).
\end{itemize}
Following this we  obtain immediately from (\ref{tw1})-(\ref{tw2})
the answer for the main singular parts of $G^{B}_{k}$ and
$H^{B}_{k}$
\begin{eqnarray}
  && G^{B}_{k}=0 , \label{g0}\\
  &&
  H^{B}_{k}=-\frac{(\ell-k)(\ell-k-1)}{(4\ell)}F^{B}_{k}
  =(-1)^{k+1}\binom{\ell-2}{k}\frac{\ell-1}{4u^{\ell+k+2}} .\label{h}
\end{eqnarray}
Checking consistency of the solutions we can insert these in
(\ref{d0})-(\ref{d}) and see that they are satisfied automatically.
So we see that for the main singular term we can put $G^{B}_{k}=0$
which will dramatically simplify the naive Ward identities.

\section{Renormalized trace}
Now we introduce a regularization for  the remaining set of
functions $F^{B}_{k}$ and $H^{B}_{k}$. Following the prescription of
\cite{RM4} we see that we can continue our $u^{-n}$ distributions
analytically away from integer $d$ and observe that regularized
distributions
\begin{eqnarray}
  &&F_{k}^{R}(u)=(-1)^{k}
 \binom{\ell}{k}\frac{1}{u^{\ell+k+d-1}} ,  \label{regu1}\\
  && H^{R}_{k}=-\frac{(\ell-k)(\ell-k-1)}{(4\ell)}F^{B}_{k}
  =(-1)^{k+1}\binom{\ell-2}{k}\frac{\ell-1}{4u^{\ell+k+d-1}}\label{regu2}
\end{eqnarray}
satisfy the regularized Ward identities (away from $d=3$)
\begin{eqnarray}
  D^{1}_{\ell}(F^{R},0)&=& (\ell-k)(u+1)
  \partial_{u}F^{R}_{k}+(k+1)u(u+2)\partial_{u}F^{R}_{k+1}\nonumber\\
  &+&(\ell-k)(\ell+d+k)F^{R}_{k}+(k+1)(\ell+d+k+1)u F^{R}_{k+1}=0
  ,\label{d1r}\quad\quad\\
  D^{2}_{\ell}(F^{R},0,H^{R})&=&(k+1)(\ell-k-1)F^{R}_{k+1}
  +2\partial_{u}H^{R}_{k}+2(k+1)H^{R}_{k+1}=0 ,\label{d2r}\quad\quad\\
T^{1}_{\ell}(F^{R},0)&=&(\ell-k)(\ell-k-1)F^{R}_{k}
+2(k+1)(\ell-k-1)(u+1)
F^{R}_{k+1}\nonumber\\&+&(k+2)(k+1)u(u+2)F^{R}_{k+2}=0
,\label{tR1} \\
  T^{2}_{\ell}(F^{R},0,H^{R})&=&(\ell-k)(\ell-k-1)F^{R}_{k}
  +2(d+2\ell-3)H^{R}_{k}=0 .\label{tR2}
\end{eqnarray}
It is the usual picture for dimensional regularization \cite{Schwim}
 when the regularized effective action is conformal invariant and a non
 zero trace arises only after subtraction of the singularities. The
 same behaviour we expect to get here.

 Indeed  we can just put in
 (\ref{d1r}), (\ref{tR2}) $d=3-\epsilon$ and declare these to be the regularized
 Ward identities. Then using the standard relation
 \begin{equation}\label{sd}
    \left[\frac{1}{u^{n-\epsilon}}\right]_{sing}=
    \frac{1}{\epsilon}\frac{(-1)^{n-1}}{(n-1)!}\delta^{(n-1)}(u) .
\end{equation}
we can split (\ref{regu1}), (\ref{regu2}) for $d=3-\epsilon$
 in singular and renormalized parts
 \begin{eqnarray}
    &&F^{R}_{k}(u)=
 \binom{\ell}{k}\left(\frac{1}{\epsilon}\frac{(-1)^{\ell+1}}
 {(\ell+k+1)!}\delta^{(\ell+k+1)}(u)+f_{k}(u)\right)
 +F^{Ren}_{k}(u),\label{ren1}\\
&&H^{R}_{k}(u)=
 \frac{\ell-1}{4}\binom{\ell-2}{k}\left(\frac{1}{\epsilon}\frac{(-1)^{\ell}}
 {(\ell+k+1)!}\delta^{(\ell+k+1)}(u)+h_{k}(u)\right)
 +H^{Ren}_{k}(u).\label{ren2}\quad\quad
\end{eqnarray}
Here we introduced also sets of finite distributions $f_{k}(u),
h_{k}(u)$ (without $\epsilon$ pole) to describe the finite
renormalization freedom. These singular parts correspond to the
local counterterms of the effective action (they are proportional to
$\delta^{(n)}(u)).$  As usual we will concentrate on the subtraction
parts
\begin{eqnarray}
    &&F^{S}_{k}(u)=
 \binom{\ell}{k}\left(\frac{1}{\epsilon}\frac{(-1)^{\ell+1}}
 {(\ell+k+1)!}\delta^{(\ell+k+1)}(u)+f_{k}(u)\right)
,\label{sub1}\\
&&H^{S}_{k}(u)=
 \frac{\ell-1}{4}\binom{\ell-2}{k}\left(\frac{1}{\epsilon}\frac{(-1)^{\ell}}
 {(\ell+k+1)!}\delta^{(\ell+k+1)}(u)+h_{k}(u)\right) ,
\label{sub2}\quad\quad
\end{eqnarray}
understanding that the renormalized correlation function formed by
$F^{Ren}_{k}(u), H^{Ren}_{k}(u)$ will  on the quantum level get the
same trace as a subtracted singular part but with opposite sign
because the regularized expression is traceless and conserved.

The general solution of our problem is the following. We have to
switch on finite (without $\epsilon$ pole) distributions
$G^{S}_{k}(u)$ and solve the conservation Ward identities
\begin{eqnarray}
  &&D^{1}_{\ell}(F^{S},G^{S})_{d=3-\epsilon}= 0 \quad ,\label{dd1}\\
  && D^{2}_{\ell}(F^{S},G^{S},H^{S})_{d=3}=0 \quad ,\label{dd2}
\end{eqnarray}
using as general ansatz short distance expansions
\begin{eqnarray}
  && f_{k}(u)=\sum^{\ell}_{p=0}g^{p}_{k}
  \frac{\delta^{(\ell+k+1-p)}(u)}{(\ell+k+1-p)!} , \label{saz1}\\
  && h_{k}(u)=\sum^{\ell}_{p=0}b^{p}_{k}
  \frac{\delta^{(\ell+k+1-p)}(u)}{(\ell+k+1-p)!} ,\label{saz2}\\
  && G^{S}_{k}(u)= \frac{\ell-1}{4}\binom{\ell-2}{k}
  \sum^{\ell-2}_{p=0}d^{p}_{k}
  \frac{\delta^{(\ell+k+1-p)}(u)}{(\ell+k+1-p)!} .\label{saz3}
\end{eqnarray}
This fixes $G^{S}_{k}(u)$ , $h_{k}(u)$ completely and a part of the
freedom in $f_{k}(u)$. Then we put this solution in the expression
for the traces $T^{1}_{\ell}(F^{S},G^{S})_{d=3},
T^{2}_{\ell}(F^{S},G^{S},H^{S})_{d=3-\epsilon}$ and removing the
remaining freedom in the coefficients $g^{p}_{k}$ obtain \emph{the
anomalous trace}
\begin{eqnarray}
  && T^{1}_{\ell}(F^{S},G^{S})_{d=3}=(-1)^{\ell+1}\ell(\ell-1)
  \binom{\ell-2}{k}\frac{2 \delta^{(\ell+k+2)}(u)}{(\ell+k+3)!} \label{an1}\\
  && T^{2}_{\ell}(F^{S},G^{S},H^{S})_{d=3-\epsilon}=(-1)^{\ell+1}\ell(\ell-1)
  \binom{\ell-2}{k}\frac{\delta^{(\ell+k+1)}(u)}{\ell(\ell+k+1)!}\label{an2}
\end{eqnarray}
The general complicated proof of this universal result will be
sketched in Appendix \textbf{B} of this article\footnote{We also
performed exact calculations with a computer program (Mathematica 5)
for the cases $\ell=2$ and $\ell=4$ with the results in agreement
with (\ref{an1}), (\ref{an2})}.

 In the main text we will only watch
the highest delta-function's derivative terms of the Ward
identities. The reason for this is the following: Anomaly is located
only in the highest derivative part of Ward identities and described
by $g^{0}_{k}$ and $b^{0}_{k}$ coefficients. The finite
distributions $G^{S}_{k}(u)$ do have no direct influence in the
expressions for anomalous traces and are necessary only for the
cancellation of the lower delta-function's derivatives in the Ward
identities and in the rigorous proof of the existence of the
solutions for (\ref{dd1}),(\ref{dd2}),(\ref{an1}) and (\ref{an2}).
This is easy to see inserting the ansatz (\ref{saz1})-(\ref{saz3})
in the Ward identities and using the relation
$u\delta^{(n)}(u)=-n\delta^{(n-1)}(u)$. As a result  we can now put
in as first approximation
\begin{eqnarray}
  && G^{S}_{k}=0 \\
  && f_{k}(u)=g^{0}_{k}\frac{\delta^{(\ell+k+1)}(u)}{(\ell+k+1)!} \\
  && h_{k}(u)=b^{0}_{k}\frac{\delta^{(\ell+k+1)}(u)}{(\ell+k+1)!}
\end{eqnarray}
and solve
\begin{eqnarray}
  &&D^{1}_{\ell}(F^{S},0)_{d=3-\epsilon}= 0 ,\label{ddd1}\\
  && D^{2}_{\ell}(F^{S},0,H^{S})_{d=3}=0 ,\label{ddd2}
\end{eqnarray}
watching only the highest delta-function's derivative terms. From
(\ref{ddd1}) and in agreement with \cite{RM4} we obtain
\begin{equation}\label{solg}
    g^{0}_{k}-g^{0}_{k+1}=\frac{1}{\ell+k+2},\quad k=0,\dots \ell-1
    .
\end{equation}
From (\ref{ddd2}) follows the recursion
\begin{equation}\label{solb}
    2\ell
    g^{0}_{k+1}-(\ell-k-2)b^{0}_{k+1}-(\ell+k+2)b^{0}_{k}=0 ,\quad  k=0,\dots
    \ell-3
\end{equation}
with the boundary condition
\begin{equation}\label{bc}
b^{0}_{\ell-2}=g^{0}_{\ell-1} ,
\end{equation}
which implies the unique solution for (\ref{solb})
\begin{equation}\label{sb}
b^{0}_{k}=g^{0}_{k}-\frac{1}{2\ell}\quad , k=0,\dots \ell-2 .
\end{equation}
Inserting this solution in the expression for the trace we obtain up
to lower order derivative terms again
\begin{eqnarray}
  && T^{1}_{\ell}(F^{S},0)_{d=3}=(-1)^{\ell+1}\ell(\ell-1)
  \binom{\ell-2}{k}\frac{2 \delta^{(\ell+k+2)}(u)}{(\ell+k+3)!} ,\\
  && T^{2}_{\ell}(F^{S},0,H^{S})_{d=3-\epsilon}=(-1)^{\ell+1}\ell(\ell-1)
  \binom{\ell-2}{k}\frac{\delta^{(\ell+k+1)}(u)}{\ell(\ell+k+1)!} .
\end{eqnarray}
Finally we can derive from (\ref{tracemap}) and (\ref{an1}),
(\ref{an2}) the following expression ($\ell$ is even)
\begin{eqnarray}
  &&TrK^{\ell}(F^{S},G^{S},H^{S})=\frac{1}{\ell(\ell-1)}\Box_{a}K^{\ell}(F^{S},G^{S},H^{S})\nonumber\\
  &&=-I^{2}_{2c}\Psi^{\ell-2}\left[\binom{\ell-2}{k}\frac{2
  \delta^{(\ell+k+2)}(u)}{(\ell+k+3)!}\right] -
  c^{2}_{2}\Psi^{\ell-2}\left[\binom{\ell-2}{k}
  \frac{\delta^{(\ell+k+1)}(u)}{\ell(\ell+k+1)!}\right] .\label{trK}
\end{eqnarray}
\section{Effective action and trace anomaly}
First we return to the singular part of the two-point function
without trace terms already calculated in \cite{RM4}
\begin{eqnarray}
  && \Pi^{\ell}_{sing}(z_{1};a|z_{2};c)=n(\ell)K^{\ell}_{sing}(F^{S},0,0)
  +\textnormal{ trace terms} ,\\
  &&K^{\ell}_{sing}(F^{S},0,0)=\Psi^{\ell}\left[\binom{\ell}{k}
  \frac{1}{\epsilon}\frac{(-1)^{\ell+1}}
 {(\ell+k+1)!}\delta^{(\ell+k+1)}(u)\right] .
\end{eqnarray}
Inserting this in (\ref{efac}) for a transversal and traceless
higher spin field $h^{(\ell)}$ and using our technique for
integration and transformation of $u$-derivatives of the delta
functions to the covariant derivatives of the four-dimensional delta
function in the general coordinate system (see \cite{RM4} for
details), we arrive at the same expression as in the previous
article
\begin{eqnarray}
  && W^{(\ell)}_{Sing}(h^{(\ell)})=\frac{1}{\epsilon}Z^{\ell}\Omega_{3}\int
   \sqrt{g}d^{4}z h^{(\ell)}_{\mu_{1}\dots\mu_{\ell}}K^{\ell}
   (\Box)h^{(\ell)\mu_{1}\dots\mu_{\ell}} ,\label{efa1}\\
  && K^{\ell}(\Box)=\left\{2\hat{D}_{\ell}
+(\ell+2)\right\}\prod^{\ell-1}_{m=1}\left[\hat{D}_{m}+\frac{\ell}{2m}\right] ,\label{kop}\\
&&\hat{D}_{n}=\frac{1}{2n}\left[\Box+2-n(n+1)\right]\,,\label{DE}\quad
Z^{\ell}=\frac{n(\ell)}{2\ell^{2}(2\ell+1)\ell!}\quad .
\end{eqnarray}
In the case of $\ell=2$ this integral should be proportional to the
integrated square of the gravitational Weyl tensor linearized in the
$AdS_{4}$ background (see \cite{BD},\cite{OP},\cite{Osborn} and ref.
there)
\begin{eqnarray}
  && C^{\mu\nu}_{\lambda\rho}(G)C^{\lambda\rho}_{\mu\nu}(G)=
  R^{\mu\nu}_{\lambda\rho}(G)R^{\lambda\rho}_{\mu\nu}(G)
  -2R^{\mu\nu}(G)R_{\mu\nu}(G)+\frac{1}{3}R(G)R(G) ,\label{c22}\\
  &&G_{\mu\nu}=g_{\mu\nu}+h^{(2)}_{\mu\nu}\quad ,\quad
  \nabla^{\mu}h^{(2)}_{\mu\nu}=h^{(2)\mu}_{\mu}=0 .
\end{eqnarray}
For traceless and transversal $h^{(2)}_{\mu\nu}$ in an $AdS_{4}$
background (we put as before $L$=1) we have
\begin{eqnarray}
  && R^{\mu\nu}_{\lambda\rho}(G)=R^{\mu\nu}_{\lambda\rho}(h^{(2)})=
  2\nabla^{[\mu}\nabla_{[\lambda}h^{(2)\nu]}_{\rho]}
  -2\delta^{[\mu}_{[\lambda}h^{(2)\nu]}_{\rho]} , \\
  && R^{\mu}_{\lambda}(h^{(2)})=\frac{1}{2}[\nabla^{\mu}\nabla_{\lambda}h^{(2)\nu}_{\nu}+\Box
  h^{(2)\mu}_{\lambda}-2\nabla^{(\mu}(\nabla\cdot h^{(2)})_{\lambda)}]
  +h^{(2)\mu}_{\lambda}-g^{\mu}_{\lambda}h^{(2)\nu}_{\nu}\nonumber\\&&\quad\quad\quad\quad=\frac{1}{2}\Box
  h^{(2)\mu}_{\lambda}+h^{(2)\mu}_{\lambda} ,\\
  &&R(h^{(2)}) =\Box
  h^{(2)\mu}_{\mu}-\nabla^{\mu}\nabla^{\nu}h^{(2)}_{\mu\nu}-3h^{(2)\mu}_{\mu}=0 ,\label{rterm}
\end{eqnarray}
and straightforward calculations lead to
\begin{equation}\label{c21}
    \int\sqrt{g}d^{4}z C^{\mu\nu}_{\lambda\rho}(h^{(2)})C^{\lambda\rho}_{\mu\nu}(h^{(2)})
    =\frac{1}{2}\int\sqrt{g}d^{4}z
    h^{(2)}_{\mu\nu}\left[\Box^{2}+6\Box+8\right]h^{(2)\mu\nu} .
\end{equation}
Then we can evaluate (\ref{kop}) for $\ell=2$ and obtain
\footnote{Restoring $L$ dependence $
K^{2}(\Box)=\frac{1}{4}\left[\Box^{2}+6\frac{1}{L^{2}}\Box+8\frac{1}{L^{4}}\right]$,
and therefore $K^{2}(\Box)\rightarrow \frac{1}{4}\Box^{2}$ when $L \rightarrow
\infty$. It means that this anomaly contribution can be determined
from the divergent part of the two point function in both the flat and the
constant curvature background \cite{OP},\cite{Osborn}.}
\begin{equation}\label{kop2}
    K^{2}(\Box)=\frac{1}{4}\left[\Box^{2}+6\Box+8\right] .
\end{equation}
So we see that
\begin{equation}\label{finfor}
     W^{(2)}_{Sing}(h^{(\ell)})=\frac{1}{\epsilon}
     \frac{Z^{2}\Omega_{3}}{2}\int\sqrt{g}d^{4}z
C^{\mu\nu}_{\lambda\rho}(h^{(2)})C^{\lambda\rho}_{\mu\nu}(h^{(2)}) .
\end{equation}

Next we compare the coefficient $\frac{Z^{2}\Omega_{3}}{2}$ with the
textbook result (see formula (6.102) in \cite{BD}). For doing this
carefully we note that our current $J^{(2)}_{\mu\nu}$ defined
according to formulas (1)-(5) of \cite{RM4} as
\begin{equation}\label{jcur}
J^{(2)}_{\mu\nu}=\frac{2}{\sqrt{g}}\frac{\delta}{\delta
h^{\mu\nu}}S^{(2)conf}_{int}=
\sigma\nabla_{\mu}\partial_{\nu}\sigma-2\partial_{\mu}\sigma\partial_{\nu}\sigma+\dots
\end{equation}
and the energy-momentum tensor of the conformal scalar in $AdS_{4}$
\begin{eqnarray}
  && T_{\mu\nu}=\frac{2}{\sqrt{g}} \frac{\delta}{\delta h^{\mu\nu}}W_{cl}(G=g+h)=
  \frac{1}{3}\sigma\nabla_{\mu}\partial_{\nu}\sigma-\frac{2}{3}
  \partial_{\mu}\sigma\partial_{\nu}\sigma+\dots ,\\
  &&W_{cl}(G=g+h)=\frac{1}{2}\int\sqrt{G}d^{4}z
  \left(G^{\mu\nu}\partial_{\mu}\sigma\partial_{\nu}\sigma-\frac{1}{6}R(G)\sigma^{2}\right)
\end{eqnarray}
are related in the following way
\begin{equation}\label{norm}
    T_{\mu\nu}=\frac{1}{3}J^{(2)}_{\mu\nu} .
\end{equation}

So we see that we can compare our spin-two  effective action with
the gravitational one after taking into account the additional
factor $[\frac{1}{3}]^{2}$. This leads to
\begin{equation}\label{anom coef}
    \frac{1}{9}\frac{Z^{2}\Omega_{3}}{2}=\frac{1}{2} \frac{1}{16\pi^{2}} \frac{1}{120}
\end{equation}
which exactly coincides with the coefficient in front of the $C^{2}$
term of the singular part of the effective action in \cite{BD} and
produces the first part of the common expression for the anomalous
trace of the energy-momentum tensor\footnote{In the literature instead
of $\alpha$ and $\beta$ one can also find other notations (e.g. $c$ and
$a$, respectively) for the coefficients of the trace anomaly \cite{Osborn}.}:
\begin{eqnarray}\label{anomaly}
    &&<T^{\mu}_{\mu}>_{sing}=\frac{2}{\sqrt{G}}G^{\mu\nu}
    \frac{\delta W_{sing}(G)}{\delta G^{\mu\nu}}=
    \frac{1}{16\pi^{2}}\left[\alpha C^{\mu\nu}_{\lambda\rho}
    (G)C^{\lambda\rho}_{\mu\nu}(G)-\beta
    E_{4}\right],\quad ,\\
   && \alpha=\frac{1}{120} , \quad\quad \beta=\frac{1}{360} .
   \nonumber
\end{eqnarray}
Note that in the conformal flat $AdS$ background
$C^{\mu\nu}_{\lambda\rho}(G=g+h)=C^{\mu\nu}_{\lambda\rho}(h)$
because $C^{\mu\nu}_{\lambda\rho}(g)=0$.

It is important that we got from a perturbative calculation only a
$C^{2}$ type $\frac{1}{\epsilon}$ pole term of the effective action
but not a topological Euler density term. This means that a
perturbative calculation with dimensional regularization of the
correlation function does not feel this kind of pole terms with
$\frac{0}{0}$ behaviour when $\epsilon$ goes to $0$ (see the
corresponding discussion in \cite{Schwim}).

Nevertheless we can extract information about this topological term
from our calculation of the non-zero finite trace of the two-point
function. First of all we will construct the effective action for
a transversal field $h^{(\ell)}$ with nonvanishing trace but vanishing double trace
\begin{eqnarray}
 &&<TrJ^{\ell}> = Tr\frac{\ell}{\sqrt{g}}\frac{\delta W^{\ell}_{sing}}{\delta
  h^{(\ell)}}=\ell\, n(\ell)TrK^{\ell}\ast_{c_{2}} h^{(\ell)}(c;z_{2})\label{trJ}
\end{eqnarray}
We will therefore decompose $TrK^{\ell}$ into three parts
\begin{equation}
 TrK^{\ell} = A + Bc_{\mu}\nabla_{2}^{\mu} + C(c_{2}^{2})^{2}
\end{equation}
so that B and C drop out after insertion into (\ref{trJ}).
Thus $A$ represents a restclass and our issue is to find an optimal representative for
the rest class of $TrK^{\ell}$. We denote rest class equivalence by $\equiv$.

The reduction of $TrK^{\ell}$ to this optimal form is achieved by a series of partial
integrations based on
\begin{equation}
 I_{2c}\delta^{(m)}(u) = c_{\mu}\nabla_{2}^{\mu}\delta^{(m-1)}(u)
\end{equation}
Any partial integration is performed employing the formulae of Appendix A.
First we  transform the first part
of (\ref{trK}) proportional to $I^{2}_{2c}$ and add to the result the
second part proportional to $c_{2}^{2}$. After $\ell-2$ times repeated partial
integrations we can render $TrK^{\ell}$ equivalent to the following intermediate form
\begin{eqnarray}
  && TrK^{\ell}\equiv -c^{2}_{2}(\ell-2)!\left\{\Psi^{\ell-2}
  \left[2\sum^{\ell-2}_{p=0}M^{p}_{k}(u)\right]\right.\nonumber\\
  &&\left.\quad\qquad\qquad +\Psi^{\ell-2}
  \left[\frac{\delta^{(\ell+k+1)}}{ k! (\ell-2-k)!\ell(\ell+k+1)!}\right]\right\} , \label{mpk}\\
  &&M^{p}_{k}(u)=\frac{(-1)^{p+1}(k+1)(u+1)\delta^{(\ell+k+1)}(u)}{k!(\ell-2-k-p)!(\ell+k+3+p)!}\nonumber\\
  &&\quad\quad\quad\quad\quad+
  \frac{(-1)^{p+1}(\ell-k-1)\delta^{(\ell+k)}(u)}{(k-1)!(\ell-1-k-p)!(\ell+k+2+p)!}
  .
\end{eqnarray}
Then using
$u\delta^{(\ell+k+1)}(u)=-(\ell+k+1)\delta^{(\ell+k)}(u)$ and the
summation rules
\begin{eqnarray}
    &&\sum^{m}_{p=0}\frac{(-1)^{p}}{(m-p)!(r+p)!}
    =\frac{1}{(m+r)(m)!(r-1)!} ,\quad\quad\label{sr}\\
    &&\sum^{m}_{p=0}\frac{(-1)^{p+1}p}{(m-p)!(r+p)!}=\frac{1}{(m+r-1
    )(m+r)(m-1)!(r-1)!} ,\quad\quad\label{sr1}
\end{eqnarray}
we obtain
\begin{eqnarray}
  && \sum^{\ell-2}_{p=0}M^{p}_{k}(u)=\frac{(\ell+1)\delta^{(\ell+k)}(u)-
  (k+1)\delta^{(\ell+k+1)}(u)}{(2\ell+1)k!(\ell-2-k)!(\ell+k+2)!} .
 \end{eqnarray}

Inserting this in (\ref{mpk}) we obtain an expression proportional to $c_{2}^{2}$ which
produces the trace of $h^{(\ell)}$.
Next we apply the same algorithm of partial integration to $\Psi^{\ell-2}$
\begin{eqnarray}
  &&
  \sum^{\ell-2}_{k=0}\phi_{k}\delta^{(m+k)}(u)I^{\ell-2-k}_{1}I^{k}_{2}
  \equiv I^{\ell-2}_{1}\delta^{(m)}(u)\sum^{\ell-2}_{k=0}(-1)^{k}\phi_{k}k!
\end{eqnarray}
and again taking into account (\ref{sr}), (\ref{sr1}) we obtain
finally
\begin{eqnarray}
  && TrK^{\ell}(u)\equiv -\frac{c^{2}_{2}I^{\ell-2}_{1}}{\ell!(2\ell+1)}
  \left[\frac{2\delta^{(\ell+1)}(u)}{2\ell-1}+\frac{\delta^{(\ell)}(u)}{\ell}\right].
\end{eqnarray}

In the next step we have to express the derivatives of the deltafunction of the chordal distance by
polynomials of the Laplacian applied to the standard deltafunction. This problem was solved by the authors in general in \cite{RM4}
\begin{equation}\label{sol2}
\delta^{(n)}(u)=(-1)^{n}\Omega_{3}\left\{2\hat{D}_{n-1}
+n\right\}\left\{\prod^{n-2}_{m=1}\hat{D}_{m}\right\}
\frac{\delta_{(4)}(z-z_{pole})}{\sqrt{g(z)}} ,
\end{equation}
with $\hat D_{m}$ as in (\ref{DE}). Moreover we need the commutation relation
\begin{equation}\label{rel1}
    \Box I_{1}^{\ell-2}\hat{\ast}_{c_{2}}h^{(\ell)}(c;z_{2}) = I_{1}^{\ell-2}
    \hat{\ast}_{c_{2}}\left\{\Box+\ell-2\right\}h^{(\ell)}(c;z_{2})
    .
\end{equation}
Using these relations and the fact that
\begin{equation}
  \lim_{u\rightarrow 0}I_{1}(a, c; u) = a^{\mu}c_{\mu}
\end{equation}
we obtain the final formula for the trace anomaly
\begin{eqnarray}
 &&<TrJ^{(\ell})> =\frac{\Omega_{3}n(\ell)}{\ell^{2}(\ell-1)\ell!(4\ell^{2}-1)}
  T^{\ell}(\Box)Trh^{(\ell)}(c;z_{2}) ,\label{T}\\
  &&T^{\ell}(\Box)=[\Box^{2}-(\ell^{2}-\ell-1)\Box-\ell(\ell^{2}-1)]
  \prod^{\ell-2}_{m=1}\left[\hat{D}_{m}+\frac{\ell-2}{2m}\right] .\label{TT}
\end{eqnarray}

Now we compare this expression with the known answer for the
$\ell=2$ case \cite{BD} with the second part of (\ref{anomaly}).
First of all the Euler density has the linear term in the linearized
AdS background
\begin{eqnarray}
  E_{4}&=& C^{\mu\nu}_{\lambda\rho}(G)C^{\lambda\rho}_{\mu\nu}(G)
  -2\left(R^{\mu\nu}(G)R_{\mu\nu}(G)-\frac{1}{3}R(G)R(G)\right)\nonumber\\
  &=&-4R(h^{(2)})=-4(\Box-3)Trh^{(2)} ,\label{e4}\\
  G_{\mu\nu}&=&g_{\mu\nu}+h^{(2)}_{\mu\nu},\quad\quad
  \nabla^{\mu}h^{(2)}_{\mu\nu}=0 .
\end{eqnarray}
Note that in the flat background $E_{4}$ starts from the quadratic
terms in $h^{(2)}$  and could be discovered only from the
three-point function \cite{OP}\footnote{In other words restoring $L$
dependence we see that the linearized Euler density is
$E_{4}=-4(\frac{1}{L^{2}}\Box-\frac{3}{L^{4}})Trh^{(2)}$ and
therefore $E_{4}\rightarrow 0$ when $L\rightarrow \infty$ } . So we
should compare our expression with the following one
\begin{eqnarray}
  && <T^{\mu}_{\mu}>=-\frac{\beta}{16\pi^{2}}4(\Box-3)Trh^{(2)}
  ,\quad  \beta=\frac{1}{360} . \label{E4}
\end{eqnarray}
Indeed inserting $\ell=2$ in (\ref{TT}) we obtain
\begin{eqnarray}
  && T^{2}(\Box)Trh^{(2)}=(\Box^{2}-\Box-6)Trh^{(2)}=\Box R +2R .
\end{eqnarray}
But the first term $\Box R$ corresponds to the regularization scheme
dependent contribution in the trace anomaly (so-called \emph{trivial
anomaly}) and can be absorbed by the appropriate choice of a finite
local counterterm $R^{2}$ \cite{BD}.

So finally we obtain the following result
\begin{eqnarray}
  &&<TrJ^{(2)}> =-\frac{1}{40\cdot 16\pi^{2}}E_{4} .\label{Tt}
\end{eqnarray}
Then remembering  the different normalization of the currents
(\ref{norm}) leading to
\\$<T T>=\frac{1}{3^{2}}<J^{(2)} J^{(2)}>$ we get
\begin{eqnarray}
  && <T^{\mu}_{\mu}>=-\frac{1}{360}\frac{1}{16\pi^{2}}E_{4} ,
\end{eqnarray}
which is in full agrement with (\ref{E4}) and \cite{BD}.

Now we return to the general $\ell$ case. We can rewrite our general
answer (\ref{T}), (\ref{TT}) using the gauge invariant object -- the
so-called "Fronsdal" operator \cite{Frons} (see for details
\cite{RM3}, \cite{RM2})
\begin{eqnarray}
\mathcal{F}(h^{(\ell )}(z;a))&=&\Box h^{(\ell )}(z;a)-(a\nabla
)\nabla ^{\mu }\frac{\partial }{\partial a^{\mu }}h^{(\ell
)}+\frac{1}{2}(a\nabla
)^{2}\Box _{a}h^{(\ell )}(z;a)\quad   \notag  \label{fe} \\
&-&\left( \ell ^{2}+\ell (d-5)-2(d-2)\right) h^{(\ell )}-a^{2}\Box
_{a}h^{(\ell)}(z;a) .  \label{F}
\end{eqnarray}
Actually we need only the trace of this object which looks very
simple for a transversal gauge field $\nabla ^{\mu }\frac{\partial
}{\partial a^{\mu }}h^{(\ell )}=0$
\begin{eqnarray}
  && Tr\mathcal{F}(h^{(\ell )}(z;a))=2\left[\Box -(\ell^{2}-1)\right]Trh^{(\ell
  )}(z;a) .
\end{eqnarray}
Introducing the new notation
\begin{equation}\label{R}
    \hat{R}^{(\ell)} Trh^{(\ell)}=\left[\Box -(\ell^{2}-1)\right]Trh^{(\ell
  )}(z;a) ,
\end{equation}
and expanding $T^{\ell}(\Box)$ (\ref{TT}) in powers of $
\hat{R}^{(\ell)}$ we obtain
\begin{eqnarray}
  &&T^{\ell}(\Box)=\frac{\hat{R}^{(\ell)}}{2^{\ell-2}(\ell-2)!}
  \prod^{\ell-2}_{m=0}\left[\hat{R}^{(\ell)}+(\ell^{2}+\ell-1)-m(m+1)\right] .\label{TTR}
\end{eqnarray}
This is a polynomial of $\ell$'th order in $\hat{R}^{(\ell)}$
without a constant term. Factorizing
\begin{eqnarray}
  && m(m+1)-(\ell^{2}+\ell-1)=(m-a_{+})(m-a_{-}) , \\
&& a_{\pm}=-\frac{1}{2}\pm \sqrt{\lambda+\frac{1}{4}} ,\quad
\lambda=\ell^{2}+\ell-1
\end{eqnarray}
we obtain the coefficient of the term linear in $\hat{R}^{(\ell)}$
\begin{equation}
    \gamma_{0}=\frac{(-a_{+})_{\ell-1}(-a_{-})_{\ell-1}}{2^{\ell-2}(\ell-2)!}
    .
\end{equation}
and as a coefficient of $\left(\hat{R}^{(\ell)}\right)^{n+1}$
\begin{eqnarray}
  && \gamma_{n}=\gamma_{0}\sum_{0\leq m_{1}< m_{2}<\dots<  m_{n}\leq \ell-2}
  \quad\prod^{n}_{i=1}\frac{1}{\ell^{2}+\ell-1-m_{i}(m_{i}+1)}.
\end{eqnarray}
Another elegant representation for $\gamma_{n}$ we can obtain
directly from (\ref{TTR}) using Taylor's expansion and replacing the
$n$'th derivative
$\left(\frac{d}{d\hat{R}}\right)^{n}(\dots)_{\hat{R}=0}$ with
$\left(\frac{d}{d\lambda}\right)^{n}(\dots)_{\hat{R}=0}$
\begin{eqnarray}
  && \gamma_{n}=\frac{1}{n!}\left(\frac{d}{d\lambda}\right)^{n}\gamma_{0}(\lambda) ,\\
  && \gamma_{0}(\lambda)=\frac{1}{2^{\ell-2}(\ell-2)!}\prod^{\ell-2}_{m=0}[\lambda-m(m+1)] ,
\end{eqnarray}
where after differentiation we have to put
$\lambda=\ell^{2}+\ell-1$.

In a similar way for the operator $K^{\ell}(\Box)$ in (\ref{kop}) we
can consider (\ref{F}) for transversal and traceless $h^{(\ell)}$
\begin{eqnarray}
  && \mathcal{F}(h^{(\ell )}(z;a))=(\Box-\ell^{2}+2\ell+2)h^{(\ell
  )}(z;a)=\hat{\mathcal{F}}^{(\ell)}h^{(\ell
  )}(z;a) ,
\end{eqnarray}
and obtain the following representation for (\ref{kop})
\begin{eqnarray}
  &&
  K^{\ell}(\Box)=\frac{\hat{\mathcal{F}}^{(\ell)}}{\ell!2^{\ell-1}}\prod^{\ell-2}_{m=0}
  \left[\hat{\mathcal{F}}^{(\ell)}+\lambda-m(m+1)\right] ,\quad
  \lambda=\ell^{2}-\ell .
\end{eqnarray}
Then expanding
\begin{eqnarray}
  && \prod^{\ell-2}_{m=0}
  \left[\hat{\mathcal{F}}^{(\ell)}+\lambda-m(m+1)\right]=
  \sum^{\ell-2}_{n=0}\hat{\gamma}_{n}(\lambda)\left(\hat{\mathcal{F}}^{(\ell)}\right)^{n}
  ,
\end{eqnarray}
we get again
\begin{eqnarray}
  && \hat{\gamma}_{n}=\frac{1}{n!}\left(\frac{d}{d\lambda}\right)^{n}\hat{\gamma}_{0}(\lambda),\\
&& \hat{\gamma}_{0}(\lambda)=\prod^{\ell-2}_{m=0}[\lambda-m(m+1)] ,
\end{eqnarray}
but with the simpler expression for $\hat{\gamma}_{0}(\lambda)$
evaluated at $\lambda=\ell^{2}-\ell$
\begin{equation}
    \hat{\gamma}_{0}(\lambda)_{\lambda=\ell^{2}-\ell}=(2\ell-2)! .
\end{equation}

\section*{Conclusions}
In this article we considered the two-point correlation function for
traceless conserved higher spin currents in $AdS_{4}$ including the
first trace terms.  We have shown that extracting the delta function
singularities we can observe the \emph{trace anomaly} in the
external higher spin gauge fields. In the particular $\ell=2$ case
our result is in full agreement with the answer for the conformal
anomaly in an external gravitational field and produces the right
anomaly numbers for both the Weyl invariant and the topological part
of anomaly. \emph{The important point here is that in the flat
background the topological part of the anomaly can not be determined from
the two point function}. This two-point function acts on transversal external
gauge fields which have either vanishing trace (case 1) or vanishing doubletrace
(case 2). In case 1 it has $\ell+1$ components corresponding to the
parts $I_{1}^{\ell-k}I_{2}^{\ell}, 0 \leq k \leq \ell$. In case 2 there
exist $2(\ell-1)$ terms corresponding to $I_{3,4}I_{1}^{\ell-2-k}I_{2}^{k},
0 \leq k \leq \ell-2$. For $\ell = 2$ the first three terms reduce to one
term, the square of the Weyl tensor with coefficient $\alpha$. The two
terms of case 2 reduce to the single topological (Euler density) term with
coefficient $\beta$. The reduction is due to equations of motion,
vanishing of traces and skipping trivial anomaly terms.

For the general spin-$\ell$ case we present
the anomaly as a polynomial of a gauge covariant differential
operator which contains information about a trivial and the topological
part of the trace anomaly. Unfortunately at the moment we have no means
to classify the possible finite local gauge invariant counterterms
constructed from the higher spin field and can neither extract the
trivial contribution in the anomaly nor define the closed form of
the topological part. But at least we know that the term linear in
$\hat{R}^{(\ell)}$  of (\ref{TTR}) should contribute to the
topological part and the corresponding numerical coefficient of this
term is presented in closed form as a function of $\ell$.
\subsection*{Acknowledgements}
\quad This work is supported by the German Volkswagenstiftung and in
part by the INTAS grant \#03-51-6346. R.M. thanks R.Mkrtchyan for
many valuable discussions.

\section*{Appendix A}
\setcounter{equation}{0}
\renewcommand{\theequation}{A.\arabic{equation}}
The Euclidian $AdS_{d+1}$ metric
\begin{equation}\label{eads}
    ds^{2}=g_{\mu \nu }(z)dz^{\mu }dz^{\nu
}=\frac{1}{(z^{0})^{2}}\delta _{\mu \nu }dz^{\mu }dz^{\nu }
\end{equation}
can be realized as an induced metric for the hypersphere defined by
the following embedding procedure in $d+2$ dimensional Minkowski
space
\begin{eqnarray}
  && X^{A}X^{B}\eta_{AB}=-X_{-1}^{2}+X_{0}^{2}+\sum^{d}_{i=1}X_{i}^{2}=-1 ,\\
  && X_{-1}(z)=\frac{1}{2}\left(\frac{1}{z_{0}}+\frac{z_{0}^{2}
  +\sum^{d}_{i=1}z^{2}_{i}}{z_{0}}\right) ,\\
  && X_{0}(z)=\frac{1}{2}\left(\frac{1}{z_{0}}-\frac{z_{0}^{2}
  +\sum^{d}_{i=1}z^{2}_{i}}{z_{0}}\right) ,\\
  && X_{i}(z)=\frac{z_{i}}{z_{0}} .
\end{eqnarray}
Using this embedding rules we can realize that the chordal distance
$\zeta(z,w)$ is just an $SO(1,d+1)$ invariant scalar product
\begin{equation}\label{gd}
    -X^{A}(z)Y^{B}(w)\eta_{AB}=\frac{1}{2z_{0}w_{0}}\left(2z_{0}w_{0}
    +\sum^{d}_{\mu=0}(z-w)^{2}_{\mu}\right)=\zeta=u+1 ,
\end{equation}
and therefore can be realized by a hyperbolic angle. Indeed we can
introduce another embedding
\begin{eqnarray}\label{hyp}
    &&X_{-1}(\Theta,\omega_{\mu})=\cosh{\Theta},\\
   && X_{\mu}(\Theta,\omega_{\mu})=\sinh{\Theta}\,\omega_{\mu}\quad,\quad\quad
    \sum^{d}_{\mu=0}\omega^{2}_{\mu}=1 ,\\
    &&ds^{2}=d\Theta^{2}+\sinh^{2}{\Theta}\, d\Omega_{d} .
\end{eqnarray}
In these coordinates the chordal distance between an arbitrary point
$X^{A}(\Theta,\Omega_{\mu})$ and the pole of the hypersphere
$Y^{A}(\Theta=0,\omega_{\mu})$ is simply
\begin{equation}\label{hd}
    \zeta= -X^{A}Y^{B}\eta_{AB}=\cosh{\Theta} .
\end{equation}
Therefore the invariant measure is expressed as
\begin{equation}\label{invv}
    \sqrt{g}d\Theta d\Omega_{d}=(\sinh\Theta)^{d}d\Theta
    d\Omega_{d}=u(u+2)^{\frac{d-1}{2}}du d\Omega_{d} .
\end{equation}
So we see that the integration measure for $d=3$ ($D=d+1=4$) will
cancel one order of $u^{-n}$ in short distance singularities and we
have to count  the singularities starting from $u^{-2}$.

 In this article we use the following rules and relations for
$u(z,z')$, $I_{1a}$, $I_{2c}$ and the bitensorial basis
$\{I_{i}\}^{4}_{i=1}$
\begin{eqnarray}
  && \Box u=(d+1)(u+1) ,\quad \nabla_{\mu}\partial_{\nu}u=g_{\mu\nu}(u+1) ,
  \quad g^{\mu\nu}\partial_{\mu}u\partial_{\nu}u=u(u+2) ,\quad\quad\quad\label{start}\\
  &&   \partial_{\mu}\partial_{\nu'}u
  \nabla^{\mu}u=u\partial_{\nu'}u ,\quad
  \partial_{\mu}\partial_{\nu'}u \nabla^{\mu}\partial_{\mu'}u
  =g_{\mu'\nu'}+\partial_{\mu'}u\partial_{\nu'}u ,\\
&&\nabla_{\mu}\partial_{\nu}\partial_{\nu'}u \nabla^{\mu}u
  =\partial_{\nu}u\partial_{\nu'}u ,\quad
  \nabla_{\mu}\partial_{\nu}\partial_{\nu'}u
  =g_{\mu\nu}\partial_{\nu'}u ,\\
&&\frac{\partial}{\partial a^{\mu}}I_{1a}\frac{\partial}{\partial
a_{\mu}}I_{1a}=u(u+2) ,\quad \frac{\partial}{\partial
a^{\mu}}I_{1}\frac{\partial}{\partial
a_{\mu}}I_{1a}=\zeta I_{2c} ,\\
&&\frac{\partial}{\partial a^{\mu}}I_{1}\frac{\partial}{\partial
a_{\mu}}I_{1}=c^{2}_{2}+ I_{2c}^{2} , \, \frac{\partial}{\partial
a^{\mu}}I_{1}\frac{\partial}{\partial a_{\mu}}I_{2}=(u+1) I_{2c}^{2}
,\,\Box_{a}I_{4}=2(d+1)c^{2}_{2} ,\\
 &&\frac{\partial}{\partial
a^{\mu}}I_{2}\frac{\partial}{\partial a_{\mu}}I_{2}=u(u+2)I_{2c}^{2}
,\quad
\Box_{a}I_{3}=2(d+1)I_{2c}^{2}+2c^{2}_{2}u(u+2) ,\\
&&\nabla^{\mu}\frac{\partial}{\partial a^{\mu}}I_{1}=(d+1)I_{2c}
,\,\nabla^{\mu}\frac{\partial}{\partial a^{\mu}}I_{2}=(d+2)(u+1)
I_{2c},\quad\nabla^{\mu} I_{1}\partial_{\mu}u=I_{2} ,\\
&&\nabla^{\mu}\frac{\partial}{\partial
a^{\mu}}I_{3}=4I_{1}I_{2c}+2(d+2)(u+1) c^{2}_{2}I_{1a}
,\quad\nabla^{\mu} I_{2}\partial_{\mu}u=2(u+1) I_{2} ,\\
&&\frac{\partial}{\partial a_{\mu}} I_{1}\partial_{\mu}u=(u+1)
I_{2c} ,\quad \frac{\partial}{\partial a_{\mu}}
I_{2}\partial_{\mu}u=u(u+2) I_{2c} ,\,\frac{\partial}{\partial
a_{\mu}}
I_{1}\nabla_{\mu} I_{1}=I_{1} I_{2c} ,\,\,\,\,\,\quad\quad\\
&&\frac{\partial}{\partial a_{\mu}} I_{1}\nabla_{\mu}
I_{2}=I_{2c}\left((u+1) I_{1}+I_{2}\right)+c^{2}_{2}I_{1a}
,\frac{\partial}{\partial a_{\mu}} I_{2}\nabla_{\mu}
I_{1}=I_{2c}I_{2} ,\\
&&\frac{\partial}{\partial a_{\mu}} I_{2}\nabla_{\mu} I_{2}=2(u+1)
I_{2c}I_{2} ,\quad \nabla^{\mu} I_{1}\nabla_{\mu}
I_{1}=a^{2}_{1}I_{2c}^{2} ,\quad \Box I_{1}=I_{1} ,\\
&&\nabla^{\mu} I_{1}\nabla_{\mu} I_{2}=I_{2}I_{1}+ a^{2}_{1}(u+1)
I_{2c}^{2} ,\quad \Box I_{2}=(d+2)I_{2}+2(u+1)
I_{1} ,\quad\\
&&\nabla^{\mu} I_{2}\nabla_{\mu} I_{2}=I_{2}^{2}+2(u+1)
I_{1}I_{2}+a^{2}_{1}I_{2c}^{2}(u+1)^{2}+c^{2}_{2}I_{1a}^{2}
,\\
&&a^{\mu}\nabla_{\mu}I_{1a}=a^{2}(u+1) ,\quad
a^{\mu}\nabla_{\mu}I_{2c}=I_{1},\quad
a^{\mu}\nabla_{\mu}I_{1}=a^{2}I_{2c},
\\&&a^{\mu}\nabla_{\mu}I_{2}=a^{2}(u+1) I_{2c}+I_{1a}I_{1},
.\label{end}
\end{eqnarray}

\section*{Appendix B}
\setcounter{equation}{0}
\renewcommand{\theequation}{B.\arabic{equation}}

We derive an algorithm to determine the finite renormalization
functions $g^{p}_{k}$ and through these the functions $h^{p}_{k},
d^{p}_{k}$ from the current conservation constraints
$D^{1}_{\ell}=0$ and $D^{2}_{\ell}=0$  (\ref{d0}), (\ref{d}) and the
tracelessness conditions $T^{1}_{\ell}\hat{=}0$,
$T^{2}_{\ell}\hat{=}0$ modulo anomalous terms already considered in
the main text. The idea is to replace $G_{k}$ and $H_{k}$ in
$D^{1}_{\ell}=0$ and $D^{2}_{\ell}=0$ by eliminating them with the
help of the two $T^{1}_{\ell}\hat{=}0$, $T^{2}_{\ell}\hat{=}0$
equations

In $D^{1}_{\ell}=0$  with
\begin{eqnarray}
  && D^{(n)}:=\frac{1}{n!}\delta^{(n)}(u) ,\label{B1}
\end{eqnarray}
and
\begin{eqnarray}
  && \Delta f_{k}:=f_{k}-f_{k+1} , \label{B2}\\
  && \Delta^{2} f_{k}=\Delta f_{k}-\Delta f_{k+1} ,\label{B3}
\end{eqnarray}
we get
\begin{eqnarray}
  && \sum_{p}D^{(\ell+k+1-p)}\left\{2\ell\left[(p+1)\Delta g^{p}_{k}+(\ell+k+4)
  \Delta g^{p+1}_{k}
   -(p+3)g^{p+1}_{k}+2(p+2)g^{p+1}_{k+1}\right]\right. \nonumber\\
  && \left. -(k+2)(\ell-k-1)\left[2\Delta
  g^{p+1}_{k+1}+\Delta^{2}g^{p}_{k}\right]-k(\ell+k+1-p)\left[2\Delta
  g^{p+1}_{k}+\Delta^{2}g^{p}_{k-1}\right]\right\}\hat{=}0 .\nonumber\\\label{B4}
\end{eqnarray}
To solve these equations we use as ansatz
\begin{eqnarray}
  && g^{\ell-q}_{k}=g^{\ell-q}_{0}+\sum^{q}_{n=1}
  \left(\sum^{n}_{r=1}Q^{(q)}_{n,r}(\ell)k^{r}\right)g^{\ell-q+n}_{0} ,\quad 0\leq
  q\leq\ell-1 ,
   \label{B5}
\end{eqnarray}
and
\begin{equation}\label{B6}
    g^{0}_{k}=g^{0}_{0}+\sum^{k}_{n=0}\frac{1}{\ell+n+1} .
\end{equation}
The functions $Q^{(q)}_{n,r}(\ell)$ are rational and can be
determined by inserting (\ref{B5}) in (\ref{B4}) recursively.
Explicit results are
\begin{eqnarray}
  && Q^{(1)}_{1,1}(\ell)=1 ,\quad
  Q^{(2)}_{2,2}(\ell)=\frac{\ell-1}{2\ell-1} ,\quad
  Q^{(2)}_{2,1}(\ell)=\frac{\ell^{2}-5\ell+1}{(\ell-1)(2\ell-1)} .\label{B7}
\end{eqnarray}
The equation $D^{1}_{\ell}=0$ is then completely exploited. The
remaining variables
\begin{equation}\nonumber
    \left\{g^{\ell-q}_{0}\right\}_{0\leq q\leq \ell}
\end{equation}
enter the system of equations derived from $D^{2}_{\ell}=0$:
\begin{equation}\label{B9}
    \sum^{r}_{q=0}M_{r,q}g^{\ell-q}_{0}=A \delta_{r,\ell} ,\quad (0\leq r
    \leq\ell) .
\end{equation}
In order that this system is consistent we need
\begin{equation}\label{B10}
    M_{0,0}=0 .
\end{equation}

In fact, the contribution to $ M_{0,0}$ is

(1) from the $F_{k+1}$ term

$+\ell(\ell-1)\binom{\ell-2}{k}$ ;

(2) from the $H'_{k}+(k+1)H_{k+1}$ term

$-\ell(\ell-1)\binom{\ell-2}{k}\left[\frac{k+2}{2\ell-1}+\frac{\ell-2-k}{2\ell-1}\right]$;

(3) from the sum over $G'_{k}, G'_{k-1}, G_{k}, G_{k-1}$  terms
after many cancellations

$-\ell(\ell-1)\binom{\ell-2}{k}\frac{\ell-1}{2\ell-1}$.

So (1), (2), (3) cancel each other. The explicit forms of
$Q^{(1)}_{1,1}$ and $Q^{(2)}_{2,2}$ (\ref{B7}) have been used in
these derivations.

There remain in this way $\ell$ equations for $\ell+1$ unknowns. We
choose $g^{0}_{0}$ as free parameter and identified this one as
renormalization parameter $\mu$ in \cite{RM4}.

\end{document}